\documentclass[10pt,conference]{IEEEtran}
\usepackage{fancyhdr}

\IEEEoverridecommandlockouts

\usepackage{cite}
\usepackage{amsmath,amssymb,amsfonts}
\usepackage{algorithmic}
\usepackage{graphicx}
\usepackage{textcomp}
\usepackage{xcolor}
\usepackage{hyperref}
\def\BibTeX{{\rm B\kern-.05em{\sc i\kern-.025em b}\kern-.08em
    T\kern-.1667em\lower.7ex\hbox{E}\kern-.125emX}}

\begin{document}

\title{Enhancing Trust in Language Model-Based Code Optimization through RLHF: A Research Design}

\author{\IEEEauthorblockN{Jingzhi Gong}
\IEEEauthorblockA{\textit{University of Leeds, UK} \\
\textit{TurinTech AI, London, UK} \\
0000-0003-4551-0701}
}

% \thanks{Submission to the Doctoral and Early Career Symposium of the Cooperative and Human Aspects of Software Engineering 2025 conference. \copyright~with the authors.}

% \IEEEnoauxwrite{\copyright~2025 IEEE. Personal use is permitted.}

\IEEEpubid{\makebox[\columnwidth]{Submission to the DECS of CHASE 2025. \copyright~with the authors.} \hspace{\columnsep}\makebox[\columnwidth]{ }}

\maketitle

\begin{abstract}
With the rapid advancement of AI, software engineering increasingly relies on AI-driven approaches, particularly language models (LMs), to enhance code performance. However, the trustworthiness and reliability of LMs remain significant challenges due to the potential for hallucinations---unreliable or incorrect responses. To fill this gap, this research aims to develop reliable, LM-powered methods for code optimization that effectively integrate human feedback. This work aligns with the broader objectives of advancing cooperative and human-centric aspects of software engineering, contributing to the development of trustworthy AI-driven solutions. 
\end{abstract}

\begin{IEEEkeywords}
Large language model, LLM, code performance optimization, genAI, AI for Software Engineering, AI4SE
\end{IEEEkeywords}

\section{Introduction}

Performance is a crucial attribute of software code \cite{DBLP:journals/corr/Gong24Deep}. With the rapid advancement of AI, software engineering increasingly relies on AI-driven approaches, particularly language models (LMs), to enhance code performance \cite{yao2024rtlrewritermethodologieslargemodels, DBLP:conf/iclr/ShypulaMZ0GYHNR24, ishida2024langpropcodeoptimizationframework}. However, a recent survey \cite{gong2025language} on LM-based code optimization highlights a significant challenge: LMs can produce hallucinations---unreliable or incorrect responses \cite{huang2024survey, sun2024autosat, peng2024perfcodegen}---which undermines the trustworthiness and reliability of LMs.

To address this gap, this research aims to develop reliable, LM-powered methods for code performance optimization that effectively integrate human feedback. The primary goals are:
\begin{enumerate}
    \item Explore various strategies for incorporating human feedback into LM-based code intelligence techniques.
    \item Design a novel code performance optimization method that leverages human feedback to enhance code performance while fostering trust in the optimization process.
\end{enumerate}

Hence, this work aligns with the Doctoral and Early Career Symposium (DECS)'s broader objectives of advancing cooperative and human-centric aspects of software engineering. The following sections detail the research design for this project.

\section{Research Design}
\begin{figure*}[!t]
  \centering
  \includegraphics[width=0.9\linewidth]{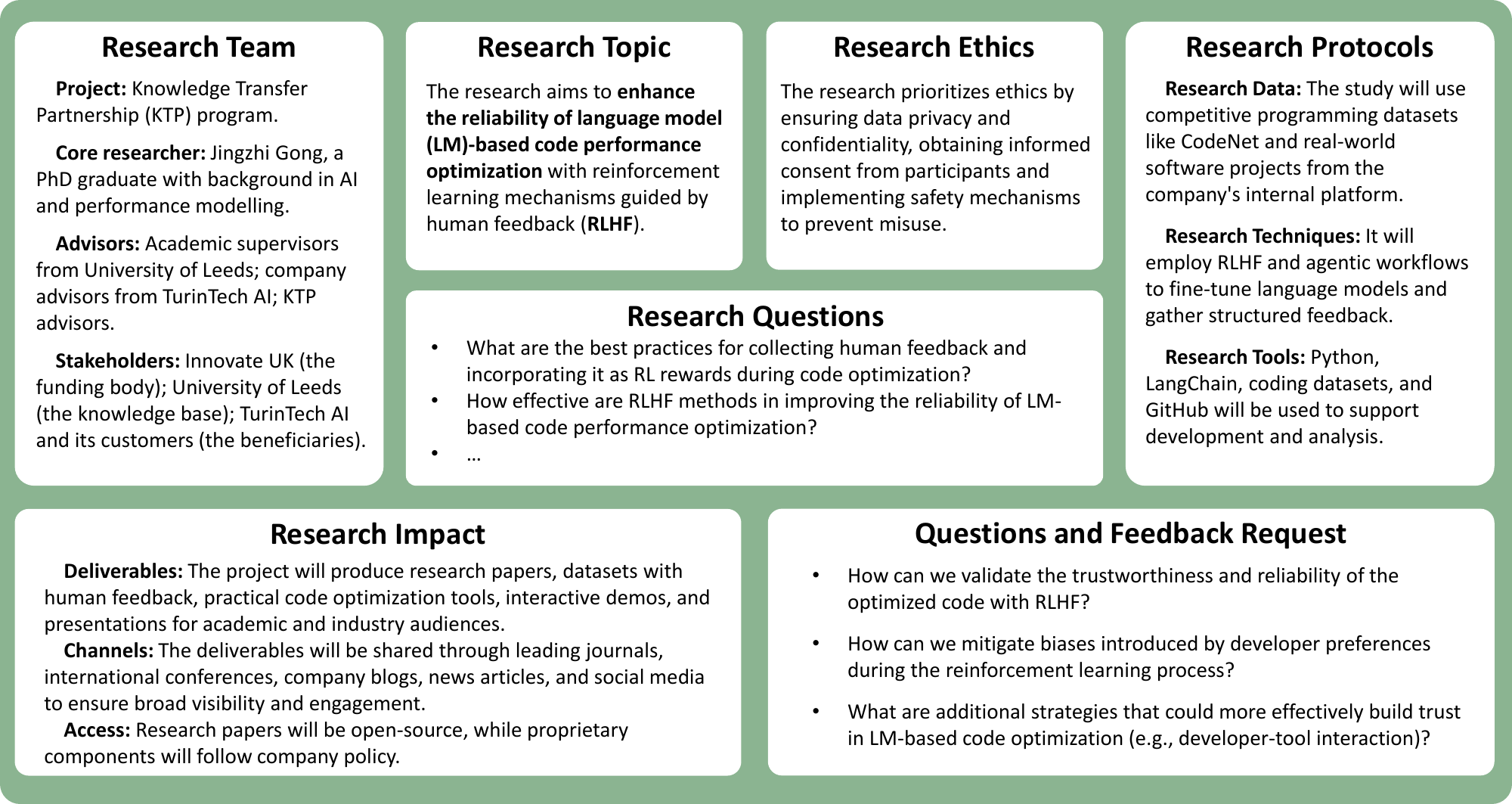}
  \caption{Overview of the research design canvas.}
  \label{fig:canvas}
\end{figure*}

This project adopts a structured methodology inspired by the Research Design Canvas \cite{hoda2024research}, as outlined in Figure~\ref{fig:canvas}.

\subsection{Research Team}
The research is conducted as part of the Knowledge Transfer Partnership (KTP) program~\cite{iukktp}, aiming at leveraging academic insights from the university to enhance the reliability of an AI-based code optimization product---Artemis \cite{turintech_artemis} at the industry partner. The research team comprises the following.

\subsubsection{Core researcher} I serve as the KTP Associate and core researcher, drawing on my expertise as a PhD graduate specializing in machine learning and software performance modeling. My role involves leading the research efforts, developing innovative methodologies, and ensuring alignment with both academic and industry goals.

\subsubsection{Advisors} The advisors are from three sides: (1) Academic supervisors from the University of Leeds provide expertise in advanced machine learning techniques and software engineering principles, offering rigorous academic guidance to ensure the research meets scholarly standards and contributes to the body of knowledge; (2) Industry supervisors from TurinTech AI, a startup specializing in the application of AI for code optimization, guide the research towards practical, real-world applications and ensure the methodologies align with industry requirements and objectives; (3) Additional oversight is offered by a KTP advisor, ensuring the project's progress aligns with its overarching objectives.

\subsubsection{Stakeholders} Key stakeholders include Innovate UK, the funding body supporting this initiative; the University of Leeds, which benefits from academic advancements from this project; and TurinTech AI along with product users, the primary beneficiaries of the research outcomes.

\subsection{Research Topic}
The research focuses on the intersection of generative AI, code optimization, and software engineering, with the goal of developing a framework that enhances the trustworthiness of LM-based code optimization with knowledge from human experts. A promising research topic therein is reinforcement learning from human feedback (RLHF) \cite{bai2022training}. Unlike traditional fine-tuning, which uses static labeled datasets to adjust model parameters, RLHF dynamically incorporates human feedback to train a reward model. This reward model guides the language model through reinforcement learning, enabling it to adapt and optimize code performance based on real-time human preferences and standards. Thereby, this approach enhances the capabilities of LMs in optimizing code performance while ensuring that human expertise and feedback play a crucial role in the process.

The topic balances interest, feasibility, and impact. It aligns with the growing interest in generative AI and its applications in software engineering, offering researchers opportunities to engage with cutting-edge techniques and methodologies. The project is feasible due to the availability of pre-trained language models, public datasets, and multidisciplinary expertise within the research team. Additionally, the integration of human feedback ensures the relevance and practicality of AI-driven solutions, offering significant impact.

\subsection{Research Ethics}
Ethics is a critical aspect of this research and is considered in several dimensions. Data privacy and confidentiality are paramount, with all data, including user feedback and proprietary code samples, being anonymized and securely stored. Informed consent is also a key consideration, ensuring that developers participating in user studies are fully informed about the purpose, benefits, and risks of their involvement. Safety mechanisms will be implemented to mitigate risks associated with the misuse of the optimization framework. Additionally, efforts will be made to ensure that the framework supports equitable outcomes and operates transparently.

\subsection{Research Questions (RQs)}
The following RQs are designed to comprehensively address the core aspects of integrating human feedback into LM-based code optimization:

\begin{enumerate}
    \item What are the best practices for collecting human feedback and incorporating it as RL rewards during code optimization?

    \item How effective are RLHF methods in improving the reliability of LM-based code optimization?

    \item How much human feedback is needed to achieve significant improvements in code reliability?

    \item What are the potential biases introduced by human feedback, and how can they be mitigated?

\end{enumerate}

Addressing these RQs ensures that the framework is not only technically sound but also aligned with the needs and preferences of developers, ultimately driving adoption and success in real-world applications.

\subsection{Research Protocols}
This section elaborates on the research data, techniques, and tools that will be used in our study to solve the above RQs.

\subsubsection{Research Data}
The research will rely on diverse datasets representing both controlled and real-world scenarios. Competitive programming datasets, such as CodeNet, will provide algorithmically rich benchmarks, making them suitable for evaluating the effectiveness of the optimization techniques. These datasets will offer structured challenges to rigorously assess the performance of the models. Additionally, real-world project datasets will be collected from the internal platform of our company to introduce practical significance, incorporating authentic examples from live projects. 

\subsubsection{Research Techniques}
The study will employ a combination of reinforcement learning methodologies and innovative workflows to achieve its objectives. RLHF will play a central role in fine-tuning language models, ensuring that the optimizations align with human preferences and real-world coding standards. To support this, agentic workflows will be used to systematically gather structured feedback from human reviewers, facilitating effective interaction between the users and the models. Furthermore, multiple language models, both open-source and proprietary, will be explored to determine the most effective frameworks for this task. 

\subsubsection{Research Tools}
The research will leverage a carefully selected set of tools to support development and analysis. Python will serve as the primary programming language for developing workflows, preprocessing data, and implementing reinforcement learning algorithms. LangChain, a specialized framework for constructing and managing language model agents, will be employed to structure the interactions between the models and the human feedback mechanisms. Additionally, supporting libraries such as NumPy and Pandas, along with open-source datasets and repositories in GitHub, will be used to facilitate data manipulation, analysis, and sharing of the research outcomes.

\subsection{Research Impact}

\subsubsection{Deliverables}
This project will deliver several impactful outputs, including research papers detailing the methodology and findings, datasets with human feedback collected for evaluation, and practical code optimization tools for the company product. Additionally, interactive demos showcasing the framework’s functionality and presentations explaining its applications will be created to engage both academic and industry audiences. 

\subsubsection{Channels}
The deliverables will be distributed through multiple channels to ensure widespread visibility and engagement. Research findings will be published in leading journals and presented at international conferences focused on AI and software engineering. Public engagement will be enhanced through company blogs, news articles, and social media posts, aimed at populating the findings to a broader audience. 

\subsubsection{Access}
Access to the research paper will be provided through open-source platforms to encourage open science. For proprietary components such as patents, access may follow the company policy to sustain the project’s impact while balancing intellectual property considerations.

% \subsection{Pilot Study}
% As part of the research process, a pilot study is planned to test and refine the proposed research design before full-scale implementation. The pilot will apply the framework in a controlled environment using publicly available datasets, complemented by real-world coding scenarios provided by industry stakeholders. This study aims to evaluate the integration of pre-trained language models with reinforcement learning mechanisms guided by human feedback, focusing on key components such as data collection protocols, NLP techniques, and analysis tools. The pilot will provide an opportunity to identify potential challenges, optimize methodologies, and validate the framework’s effectiveness in addressing diverse code optimization tasks. Feedback from academic supervisors, industry partners, and end users will play a crucial role in refining the research design, ensuring it is both methodologically sound and aligned with real-world applications. This iterative process will serve as a foundation for scaling the research to broader and more complex contexts.

\section{Questions and Feedback Request}
To advance this work, I seek feedback from DECS participants on several critical aspects of the research. 

First, I would appreciate guidance on ensuring trustworthiness and reliability in the optimization outcomes generated by the framework. Specifically, \emph{how can we validate that the optimized codes with RLHF are more trustful and reliable?} I am interested in exploring methods to evaluate and visualize the reliability of the optimization process, ensuring that developers can understand and trust the changes made by the AI.

Additionally, I welcome suggestions on approaches to mitigating biases introduced by developer preferences during the reinforcement learning process. For example, \emph{how can we account for varying preferences for code style, efficiency, or maintainability while ensuring that the model learns robust, generalized optimization strategies?} Ideas for integrating diverse perspectives into the training process to minimize the risk of overfitting to specific styles or preferences would be particularly helpful.

Finally, given the multiple approaches to enhancing cooperative and human-centric aspects in LM-based code performance optimization tools---such as using developer feedback as input, improving developer-tool interaction, and assessing the tool’s impact on coding practices---\emph{what are additional strategies that could more effectively build trust in LM-based code optimization?} Insights on how to balance these methods or adopt alternative approaches would be invaluable.

Your feedback will be instrumental in refining the research approach and ensuring that the developed framework meets the practical needs of the software engineering community. Thank you for your valuable insights and contributions.

\section{Conclusion}
This research design addresses the critical challenge of trustworthiness and reliability in LM-based code performance optimization by integrating human feedback into the optimization process. By exploring various strategies for incorporating human feedback and designing a novel code optimization method, this work will enhance the performance optimization capabilities of language models while ensuring that human expertise plays a crucial role. Questions to discuss at DECS include methods for validating the optimized codes, mitigating biases introduced by developer preferences, and identifying alternative approaches to foster trust and acceptance within the developer community. 

\bibliographystyle{IEEEtran}
\bibliography{references}

\end{document}